Characterization of microscopic deformation through two-point spatial correlation function

Guan-Rong Huang[1,2], Bin Wu[3], Yangyang Wang[4*], and Wei-Ren Chen[3#]

[1] *Physics Division, National Center for Theoretical Sciences, Hsinchu 30013, Taiwan*

[2] *Shull Wollan Center, the University of Tennessee and Oak Ridge National Laboratory, Oak Ridge, TN 37831, USA*

[3] *Biology and Soft Matter Division, Oak Ridge National Laboratory, Oak Ridge, TN 37831, USA*

[4] *Center for Nanophase Materials Sciences, Oak Ridge National Laboratory, Oak Ridge, TN 37831, USA*

[*] Correspondence: wangy@ornl.gov; Tel: +1-865-241-8495.

[#] Correspondence: chenw@ornl.gov; Tel: +1-865-574-7979.

Abstract

The molecular rearrangements of most fluids under flow and deformation do not directly follow the macroscopic strain field. In this work, we describe a phenomenological method for characterizing such non-affine deformation via the anisotropic pair distribution function (PDF). We demonstrate now the microscopic strain can be calculated in both simple shear and uniaxial extension, by perturbation expansion of anisotropic PDF in terms of real spherical harmonics. Our results, given in the real as well as the reciprocal space, can be applied in spectrum analysis of small-angle scattering experiments and non-equilibrium molecular dynamics simulations of soft matter under flow.

I.  INTRODUCTION

Understanding the flow and deformation behavior of complex fluids is an important subject in soft matter research. When a macroscopic strain or stress is applied, the atoms in a complex fluid are displaced from their equilibrium positions. It has been widely recognized that such molecular rearrangements are complicated in nature and generally do not directly follow the macroscopic strain [1-23]. How to characterize the microscopic deformation of a fluid in the non-equilibrium state is, therefore, a central problem for experimentalists and theorists alike.

As the pair distribution function (PDF) describes the local structure of a liquid, it offers a natural starting point for addressing the aforementioned challenge. Following the perturbation expansion strategy first outlined by Irving and Kirkwood [24], the anisotropic pair distribution function $g(\boldsymbol{r})$ under flow and deformation has been expressed in terms of spherical harmonics by a number of researchers [1-13, 19-23, 25-28]. The use of spherical harmonics arises because of the symmetry defined by a specific type of flow (e.g. shear or uniaxial extension). In particular, it has been shown that for weak shear flow [4, 8, 11, 20],

$$g(\boldsymbol{r}) - g(r) \propto \text{Wi} \cdot Y_2^{-2}(\theta,\phi) r \frac{dg(r)}{dr}, \quad (1)$$

where $g(r)$ is the isotropic PDF (radial distribution function) in the equilibrium state, $Y_2^{-2}(\theta, \phi)$ is the real spherical harmonic function for $l = 2$ and $m = -2$, and Wi, the so-called Weissenberg number, is the product of shear rate $\dot{\gamma}$ and some characteristic relaxation time $\tau$. Since Wi gives a measure of strain, this formula in principle allows one to determine the microscopic deformation by analyzing the anisotropic component of $g(\boldsymbol{r})$, from either scattering experiments (via Fourier transform) or computer simulation [1-13, 18-23].

However, recent experimental and computational studies have demonstrated that the microscopic strain in complex fluids undergoing shear flow is generally not only non-affine, but also dependent on the molecular position [14-23]. In other words, Eq. (1), which assumes a uniform strain in space, does not apply to real fluids. In this work, we seek to generalize Eq. (1) by considering spatially varying strain. The resulting formulas, given in both real and reciprocal spaces, serve as a useful tool for extracting microscopic deformation from small-angle scattering experiments and non-equilibrium molecular dynamics simulations of flowing soft matter.

This paper is organized as follows: First, we demonstrate, with the case of shear flow, how to incorporate spatially non-uniform strain in the Irving-Kirkwood perturbation expansion of the pair distribution function. We then proceed to the case of uniaxial extension, which is another common flow geometry. Lastly, we test the validity of the derived formulas by examining the affine deformation model of Gaussian chains and non-equilibrium molecular dynamics simulations of interacting particles.

## II. PERTURBATION EXPANSION FOR SHEAR FLOW

### A. Perturbation Expansion with Spatially Varying Strain

The pair distribution function $g(\mathbf{r})$ is generally defined as:

$$\rho g(\mathbf{r}) = \frac{1}{N} \sum_{i,j} \langle \delta(\mathbf{r} - \mathbf{r}_{ij}) \rangle, \tag{2}$$

where $\rho$ is the mean particle density, $N$ is the total number of particles, $\mathbf{r}_{ij} = \mathbf{r}_j - \mathbf{r}_i$ with $\mathbf{r}_i$ and $\mathbf{r}_j$ being the position vectors of the $i$-th and $j$-th particles, respectively, $\delta$ is the Dirac delta function, and $\langle \cdots \rangle$ in our current context stands for averaging in the configuration space. Following the perturbation expansion approach by Irving and Kirkwood [24], we seek to derive the anisotropic PDF $g(\mathbf{r})$ from the isotropic PDF $g(r)$ in the quiescent state. Since we are interested in the microscopic deformation, we will explicitly use strain $\gamma$ to describe the molecular displacements, instead of the Weissenberg number Wi, which is the traditional language Hanley, Evans, Hess and coworkers adopted when considering non-equilibrium fluids [4-12]. Moreover, we allow the microscopic strain to be dependent on the molecular position and further assume that it is a function of only the inter-particle distance $r_{ij} = |\bm{r}_{ij}|$. We can thus define a local deformation gradient tensor $\bm{E}$ for each pair of particles $i$ and $j$,

$$\bm{E} = \begin{pmatrix} 1 & \gamma(r_{ij}) & 0 \\ 0 & 1 & 0 \\ 0 & 0 & 1 \end{pmatrix}. \tag{3}$$

which describes the transformation of $\bm{r}_{ij}$ under flow: $\bm{r}'_{ij} = \bm{E} \cdot \bm{r}_{ij}$. Hence, the anisotropic PDF in the non-equilibrium state can be expressed as [24, 29]:

$$\rho g(\bm{r}) = \frac{1}{N}\sum_{i,j}\langle\delta(\bm{r}-\bm{r}'_{ij})\rangle = \frac{1}{N}\sum_{i,j}\langle\delta(\bm{r}-\bm{r}_{ij}-(\bm{E}-\bm{I})\cdot\bm{r}_{ij})\rangle, \tag{4}$$

where $\bm{I}$ is the isotropic tensor. Perturbation expansion of $\delta(\bm{r}-\bm{r}_{ij}-(\bm{E}-\bm{I})\cdot\bm{r}_{ij})$ gives:

$$\delta(\bm{r}-\bm{r}'_{ij}) - \delta(\bm{r}-\bm{r}_{ij}) = -\left\{\left[(\bm{E}-\bm{I})\cdot\bm{r}_{ij}\right]\cdot\nabla_r\right\}\delta(\bm{r}-\bm{r}_{ij}) + \frac{1}{2}\left\{\left[(\bm{E}-\bm{I})\cdot\bm{r}_{ij}\right]\cdot\nabla_r\right\}^2 \delta(\bm{r}-\bm{r}_{ij}) - \cdots \\ + \frac{1}{n!}\left\{-\left[(\bm{E}-\bm{I})\cdot\bm{r}_{ij}\right]\cdot\nabla_r\right\}^n \delta(\bm{r}-\bm{r}_{ij}) + \cdots \tag{5}$$

In the original derivation of Irving and Kirkwood, they consider a constant small perturbation $\alpha\bm{R}$ at each particle separation [24]. Under this condition, Eq. (5) leads to an elegant expansion formula for the pair distribution function. In the case of shear, when $\bm{E}$ does not depend on the molecular position, i.e., $\gamma(r_{ij})$ is a constant, Eq. (5) coincides with the Irving-Kirkwood formula that links the non-equilibrium PDF $g(\bm{r})$ to the PDF $g(r)$ in the quiescent state:

$$g(\bm{r}) - g(r) = -\left\{\left[(\bm{E}-\bm{I})\cdot\bm{r}\right]\cdot\nabla_r\right\}g(r) + \frac{1}{2}\left\{\left[(\bm{E}-\bm{I})\cdot\bm{r}\right]\cdot\nabla_r\right\}^2 g(r) - \cdots \\ + \frac{1}{n!}\left\{-\left[(\bm{E}-\bm{I})\cdot\bm{r}\right]\cdot\nabla_r\right\}^n g(r) + \cdots \tag{6}$$

However, when the molecular strain is spatially dependent, Eq. (6) is no longer valid. We need to re-examine the derivation leading up to Eq. (6). Starting from Eqs. (4) and (5), we see that:

$$g(\boldsymbol{r}) - g(r) = \frac{1}{\rho N} \sum_{i,j} \int d\boldsymbol{r}_1 \int d\boldsymbol{r}_2 \ldots \int d\boldsymbol{r}_N \psi(\boldsymbol{r}_1, \boldsymbol{r}_2, \ldots \boldsymbol{r}_N) \sum_{n=1}^{\infty} \frac{1}{n!} \left\{ -\left[(\mathbf{E}-\mathbf{I}) \cdot \boldsymbol{r}_{ij}\right] \cdot \nabla_r \right\}^n \delta(\boldsymbol{r} - \boldsymbol{r}_{ij}), \quad (7)$$

where we explicitly write out the configuration space averaging as the integral with the distribution function $\psi(\boldsymbol{r}_1, \boldsymbol{r}_2, \cdots \boldsymbol{r}_N)$. Using the following identities for the Dirac delta function [24]:

$$\frac{\partial^n}{\partial x^n} \delta(x - x_{ij}) = (-1)^n \frac{\partial^n}{\partial x_{ij}^n} \delta(x - x_{ij}), \quad (8)$$

$$\int \left[\frac{\partial^n}{\partial x^n} \delta(x)\right] f(x) dx = (-1)^n \int \delta(x) \left[\frac{\partial^n}{\partial x^n} f(x)\right] dx, \quad (9)$$

we can show that for each integral in Eq. (7):

$$\int d\boldsymbol{r}_1 \int d\boldsymbol{r}_2 \ldots \int d\boldsymbol{r}_N \psi(\boldsymbol{r}_1, \boldsymbol{r}_2, \ldots \boldsymbol{r}_N) \frac{1}{n!} \left\{ -\left[(\mathbf{E}-\mathbf{I}) \cdot \boldsymbol{r}_{ij}\right] \cdot \nabla_r \right\}^n \delta(\boldsymbol{r} - \boldsymbol{r}_{ij})$$

$$= \frac{1}{n!} (-1)^n \int d\boldsymbol{r}_1 \int d\boldsymbol{r}_2 \ldots \int d\boldsymbol{r}_N \psi(\boldsymbol{r}_1, \boldsymbol{r}_2, \ldots \boldsymbol{r}_N) \left[\gamma(r_{ij}) y_{ij} \frac{\partial}{\partial x}\right]^n \delta(\boldsymbol{r} - \boldsymbol{r}_{ij})$$

$$= \frac{1}{n!} \int d\boldsymbol{r}_1 \int d\boldsymbol{r}_2 \ldots \int d\boldsymbol{r}_N \psi(\boldsymbol{r}_1, \boldsymbol{r}_2, \ldots \boldsymbol{r}_N) \left[\gamma(r_{ij}) y_{ij} \frac{\partial}{\partial x_{ij}}\right]^n \delta(\boldsymbol{r} - \boldsymbol{r}_{ij}) \quad , \quad (10)$$

$$= \frac{1}{n!} (-1)^n \int d\boldsymbol{r}_1 \int d\boldsymbol{r}_2 \ldots \int d\boldsymbol{r}_N \delta(\boldsymbol{r} - \boldsymbol{r}_{ij}) \left(y_{ij} \frac{\partial}{\partial x_{ij}}\right)^n \left[\gamma^n(r_{ij}) \psi(\boldsymbol{r}_1, \boldsymbol{r}_2, \ldots \boldsymbol{r}_N)\right]$$

$$= \frac{1}{n!} (-1)^n y^n \frac{\partial^n}{\partial x^n} \left[\gamma^n(r) \psi_{ij}(\boldsymbol{r})\right]$$

where $\psi_{ij}(\boldsymbol{r})$ is the distribution function of $i$-th and $j$-th particles in the *quiescent* state, defined as:

$$\psi_{ij}(\boldsymbol{r}) \equiv \int d\boldsymbol{r}_1 \int d\boldsymbol{r}_2 \ldots \int d\boldsymbol{r}_N \delta(\boldsymbol{r} - \boldsymbol{r}_{ij}) \psi(\boldsymbol{r}_1, \boldsymbol{r}_2, \ldots \boldsymbol{r}_N) = \rho g(r). \quad (11)$$

Therefore, the final general expression for $g(\boldsymbol{r}) - g(r)$ in shear flow is:

$$g(\boldsymbol{r}) - g(r) = \sum_{n=1}^{\infty} \frac{1}{n!} (-1)^n y^n \frac{\partial^n}{\partial x^n} \left[\gamma^n(r) g(r)\right]. \quad (12)$$

Notice that when $\gamma(r)$ is a constant, Eq. (12) reduces to the classical Irving-Kirkwood formula [Eq. (6)]. Mathematically, it is convenient to set the following change of variables: $\gamma(r) = \lambda u(r)$, where the constant $\lambda$ is the largest microscopic strain of the system. Eq. (12) can thus be rewritten as:

$$g(\boldsymbol{r}) = g(r) + \sum_{n=1}^{\infty} \frac{1}{n!}(-1)^n \lambda^n y^n \frac{\partial^n}{\partial x^n}\left[u^n(r)g(r)\right]. \tag{13}$$

The equation expresses the anisotropic PDF in terms of a power series of $\lambda$ and derivatives of $g(r)$, with the zeroth order term being the quiescent $g(r)$.

B. First-Order Expansion

In order to apply Eq. (12) (or equivalently Eq. (13)) to extract the spatially dependent strain $\gamma(r)$ from small-angle scattering experiments or non-equilibrium molecular dynamics simulations, it is helpful to expand the anisotropic PDF $g(\boldsymbol{r})$ and structure factor $S(\boldsymbol{Q})$ as a linear combination of real spherical harmonics $Y_l^m(\theta, \phi)$:

$$g(\boldsymbol{r}) = \sum_{l,m} g_l^m(r) Y_l^m(\theta,\phi), \tag{14}$$

$$S(\boldsymbol{Q}) = \sum_{l,m} S_l^m(Q) Y_l^m(\theta,\phi). \tag{15}$$

The real and reciprocal space expansion coefficients are related through the spherical Bessel transformation:

$$g_l^m(r) = \frac{i^l}{2\pi^2 \rho} \int S_l^m(Q) j_l(Qr) Q^2 dQ, \tag{16}$$

$$S_l^m(Q) = 4\pi i^l \rho \int g_l^m(r) j_l(Qr) r^2 dr. \tag{17}$$

where $j_l(x)$ is the spherical Bessel function of the first kind.

Using the real spherical harmonic expansion approach, we avoid dealing directly with the vector-variable functions $g(\boldsymbol{r})$ and $S(\boldsymbol{Q})$. Instead, all the analyses center around the expansion coefficients $g_l^m(r)$ and $S_l^m(Q)$. In computer simulation, where the information about positions of all particles is readily available, $g_l^m(r)$ and $S_l^m(Q)$ can be straightforwardly computed:

$$g_l^m(r) = \frac{1}{4\pi} \int_0^\pi d\theta \int_0^{2\pi} d\phi\, g(r,\theta,\phi) Y_l^m(\theta,\phi) \sin\theta, \tag{18}$$

$$S_l^m(Q) = \frac{1}{4\pi} \int_0^\pi d\theta \int_0^{2\pi} d\phi\, S(Q,\theta,\phi) Y_l^m(\theta,\phi) \sin\theta. \tag{19}$$

In the case of small-angle scattering experiments, one probes the cross-sections of $S(\mathbf{Q})$ on certain planes. The method for obtaining $S_l^m(Q)$ from the two-dimensional anisotropic spectra has been discussed in detail elsewhere [30].

With the expansion coefficients $g_l^m(r)$ and $S_l^m(Q)$ in hand, we can then proceed to the analysis of the microscopic deformation by using Eq. (13). The first order term in Eq. (13) is:

$$-\lambda y \frac{\partial}{\partial x}[u(r)g(r)] = -\sin^2\theta \cos\phi \sin\phi\, r \frac{d}{dr}[\gamma(r)g(r)] = -\frac{1}{\sqrt{15}} Y_2^{-2}(\theta,\phi) r \frac{d}{dr}[\gamma(r)g(r)]. \tag{20}$$

Therefore, it is easy to see from Eq. (18) that

$$g_2^{-2}(r) = -\frac{1}{\sqrt{15}} r \frac{d}{dr}[\gamma(r)g(r)]. \tag{21}$$

Solving Eq. (21) yields:

$$\gamma(r) = \gamma(D_0) \frac{g(D_0)}{g(r)} - \frac{\sqrt{15}}{g(r)} \int_{D_0}^r \frac{g_2^{-2}(r')}{r'} dr', \tag{22}$$

where $D_0$ is the diameter of the particle. Eqs. (21) and (22) can be used as the working formulas for characterizing the spatially dependent microscopic strain $\gamma(r)$, particularly for computer simulations. For small-angle scattering experiments, it is more convenient to perform analysis in the reciprocal space. From Eqs. (17) and (21), one finds that:

$$S_2^{-2}(Q) = \frac{4\pi\rho}{\sqrt{15}} \int_0^\infty j_2(Qr) r^3 \frac{d}{dr}[\gamma(r)g(r)] dr = \frac{4\pi\rho}{\sqrt{15}} \int_0^\infty \gamma(r)g(r) r \left[ r\cos(Qr) - \frac{\sin(Qr)}{Q} \right] dr. \tag{23}$$

In principle, one can experimentally determine the isotropic PDF $g(r)$ by Fourier transform of the structure factor $S(Q)$ in the quiescent state. $S_2^{-2}(Q)$ can be obtained from analyzing the anisotropic two-dimensional spectra of the velocity-velocity-gradient and velocity-vorticity planes, i.e., the so-called 1-2 and 1-3 planes [22, 30]. Therefore, $\gamma(r)$ can be found by fitting the integral equation [Eq. (23)].

C. Second-Order Expansion

Now let us consider the second-order term in Eq. (13):

$$\frac{1}{2}\lambda^2 y^2 \frac{\partial^2}{\partial x^2}\left[u^2(r)g(r)\right] \\ = \frac{1}{2}\sin^2\theta\sin^2\phi\left\{(1-\sin^2\theta\cos^2\phi)r\frac{d}{dr}\left[\gamma^2(r)g(r)\right]+\sin^2\theta\cos^2\phi r^2\frac{d^2}{dr^2}\left[\gamma^2(r)g(r)\right]\right\}. \quad (24)$$

Combining this result with the analysis in Section II.A and B, it can be shown (by using Eqs. (13) and (18)) that the expansion up to the second order in $\lambda$ involves only the following real spherical harmonics: $Y_0^0$, $Y_2^{-2}$, $Y_2^0$, $Y_4^0$, and $Y_4^4$, with

$$g_0^0(r) - g(r) = \frac{2}{15}r\frac{d}{dr}\left[\gamma^2(r)g(r)\right] + \frac{1}{30}r^2\frac{d^2}{dr^2}\left[\gamma^2(r)g(r)\right], \quad (25)$$

$$g_2^0(r) = -\frac{\sqrt{5}}{42}r\frac{d}{dr}\left[\gamma^2(r)g(r)\right] - \frac{1}{21\sqrt{5}}r^2\frac{d^2}{dr^2}\left[\gamma^2(r)g(r)\right], \quad (26)$$

$$g_4^0(r) = -\frac{1}{210}r\frac{d}{dr}\left[\gamma^2(r)g(r)\right] + \frac{1}{210}r^2\frac{d^2}{dr^2}\left[\gamma^2(r)g(r)\right], \quad (27)$$

$$g_4^4(r) = \frac{1}{6\sqrt{35}}r\frac{d}{dr}\left[\gamma^2(r)g(r)\right] - \frac{1}{6\sqrt{35}}r^2\frac{d^2}{dr^2}\left[\gamma^2(r)g(r)\right]. \quad (28)$$

Note that the expression for $g_2^{-2}$ is already given by Eq. (21), as the second-order term in $\lambda$ does not contribute to this symmetry.

Eqs (25) through (28) can also be used, in addition to Eq. (21), for analysis of $\gamma(r)$ for the expansion to the second order. Focusing on $g_0^0(r)$ and defining $H(r) \equiv \gamma^2(r)g(r)$, we have:

$$r^4 \frac{d^2H(r)}{dr^2} + 4r^3 \frac{dH(r)}{dr} = 30r^2 \left[ g_0^0(r) - g(r) \right],$$

$$\frac{d}{dr}\left[ r^4 \frac{dH(r)}{dr} \right] = 30r^2 \left[ g_0^0(r) - g(r) \right],$$

$$\frac{dH(r)}{dr} = \frac{c}{r^4} + \frac{30}{r^4} \int_{D_0}^{r} r'^2 \left[ g_0^0(r') - g(r) \right] dr', \tag{29}$$

where $c = D_0^4 \left.\frac{dH(r)}{dr}\right|_{r=D_0}$. Carrying through the integration, we obtain the expression for $\gamma^2(r)$:

$$\gamma^2(r) = \frac{3\gamma^2(D_0)D_0^3 g(D_0) + c}{3D_0^3 g(r)} - \frac{c}{3r^3 g(r)} + \frac{30}{g(r)} \int_{D_0}^{r} \frac{1}{t^4} dt \int_{D_0}^{t} r'^2 \left[ g_0^0(r') - g(r') \right] dr'. \tag{30}$$

Similarly, applying Eq. (17) to Eq. (25), we have:

$$S_0^0(Q) - S(Q)$$

$$= 4\pi\rho \int \left[ g_0^0(r) - g(r) \right] j_0(Qr) r^2 dr$$

$$= \frac{2\pi\rho}{15} \int_0^\infty \frac{d}{dr}\left[ r^4 \frac{dH(r)}{dr} \right] j_0(Qr) dr = \frac{2\pi\rho}{15} Q \int_0^\infty r^4 j_1(Qr) \frac{d}{dr}\left[ \gamma^2(r) g(r) \right] dr \tag{31}$$

$$= \frac{2\pi\rho}{15} Q \int_0^\infty \gamma^2(r) g(r) \left[ \frac{2r^2}{Q} \cos(Qr) - \left( \frac{2r}{Q^2} + r^3 \right) \sin(Qr) \right] dr$$

As in the case of Eq. (23), Eq. (31) can be used to determine $\gamma^2(r)$ from small-angle scattering experiments as well as computer simulations.

D. A Few Remarks

At this point, it is helpful to make a few remarks about our derivations so far. First, when the microscopic strain is spatially uniform, i.e., $\gamma(r)$ is a constant, Eqs. (21), (23), (25), and (31) can be simplified, yielding elegant expression for $\gamma$ in both real and reciprocal spaces:

$$\gamma = -\frac{\sqrt{15} g_2^{-2}(r)}{r \frac{dg(r)}{dr}}, \tag{32}$$

$$\gamma = \frac{\sqrt{15} S_2^{-2}}{Q \frac{dS(Q)}{dQ}}, \tag{33}$$

$$\gamma^2 = \frac{30 \left[ g_0^0(r) - g(r) \right]}{r^2 \frac{d^2 g(r)}{dr^2} + 4r \frac{dg(r)}{dr}}, \tag{34}$$

$$\gamma^2 = \frac{30 \left[ S_0^0(Q) - S(Q) \right]}{Q^2 \frac{d^2 S(Q)}{dQ^2} + 4Q \frac{dS(Q)}{dQ}}. \tag{35}$$

In particular, Eqs. (32) and (33) have been found by the previous studies [4, 8, 11, 20].

Second, as revealed by Eq. (25), the approximation $g_0^0(r) = g(r)$ is valid up to the first order expansion in $\lambda$. Additionally, with the condition:

$$\frac{d\gamma(r)}{dr} g(r) \ll \gamma(r) \frac{dg(r)}{dr} \Rightarrow \frac{d \ln \gamma(r)}{dr} \ll \frac{d \ln g(r)}{dr}, \tag{36}$$

Eq. (21) can be rewritten as:

$$\gamma(r) = -\frac{\sqrt{15} g_2^{-2}}{r \frac{dg(r)}{dr}} \approx -\frac{\sqrt{15} g_2^{-2}}{r \frac{dg_0^0(r)}{dr}}, \tag{37}$$

which is the formula used in a previous study [20]. However, Eq. (21) in general should work better than Eq. (37), as it does not require these additional conditions.

III. PERTURBATION EXPANSION FOR UNIAXIAL EXTENSIONAL FLOW

A. Perturbation Expansion with Spatially Varying Strain

Having made the analysis with simple shear, we now turn our attention to uniaxial extension, which is another important flow geometry in rheology. Once again, we will apply the Irving-Kirkwood perturbation expansion technique, with the consideration of spatially varying microscopic strain. Denoting the microscopic stretching ratio and engineering strain by $\lambda(r_{ij})$ and $\varepsilon(r_{ij})$, respectively, we have:

$$\mathbf{r}'_{ij} = \mathbf{E} \cdot \mathbf{r}_{ij} = \begin{pmatrix} \frac{1}{\sqrt{\lambda(r_{ij})}} & 0 & 0 \\ 0 & \frac{1}{\sqrt{\lambda(r_{ij})}} & 0 \\ 0 & 0 & \lambda(r_{ij}) \end{pmatrix} \cdot \mathbf{r}_{ij} \approx \begin{pmatrix} 1 - \frac{\varepsilon(r_{ij})}{2} & 0 & 0 \\ 0 & 1 - \frac{\varepsilon(r_{ij})}{2} & 0 \\ 0 & 0 & 1 + \varepsilon(r_{ij}) \end{pmatrix} \cdot \mathbf{r}_{ij}. \quad (38)$$

Similarly, the starting point for the derivation is Eq. (7). For each integral in the summation:

$$\int d\mathbf{r}_1 \int d\mathbf{r}_2 \ldots \int d\mathbf{r}_N \psi(\mathbf{r}_1, \mathbf{r}_2, \ldots \mathbf{r}_N) \frac{1}{n!} \left\{ -\left[ (\mathbf{E} - \mathbf{I}) \cdot \mathbf{r}_{ij} \right] \cdot \nabla_r \right\}^n \delta(\mathbf{r} - \mathbf{r}_{ij})$$

$$= \frac{1}{n!}(-1)^n \int d\mathbf{r}_1 \int d\mathbf{r}_2 \ldots \int d\mathbf{r}_N \psi(\mathbf{r}_1, \mathbf{r}_2, \ldots \mathbf{r}_N) \left[ \varepsilon(r_{ij}) \left( -\frac{1}{2} x_{ij} \frac{\partial}{\partial x} - \frac{1}{2} y_{ij} \frac{\partial}{\partial y} + z_{ij} \frac{\partial}{\partial z} \right) \right]^n \delta(\mathbf{r} - \mathbf{r}_{ij})$$

$$= \frac{1}{n!} \int d\mathbf{r}_1 \int d\mathbf{r}_2 \ldots \int d\mathbf{r}_N \psi(\mathbf{r}_1, \mathbf{r}_2, \ldots \mathbf{r}_N) \left[ \varepsilon(r_{ij}) \left( -\frac{1}{2} x_{ij} \frac{\partial}{\partial x_{ij}} - \frac{1}{2} y_{ij} \frac{\partial}{\partial y_{ij}} + z_{ij} \frac{\partial}{\partial z_{ij}} \right) \right]^n \delta(\mathbf{r} - \mathbf{r}_{ij})$$

$$= \frac{1}{n!}(-1)^{n-1} \left( \frac{1}{2} \right)^n \int d\mathbf{r}_1 \int d\mathbf{r}_2 \ldots \int d\mathbf{r}_N \delta(\mathbf{r} - \mathbf{r}_{ij})$$

$$\times \sum_{\alpha+\beta+\gamma=n} \frac{n!}{\alpha!\beta!\gamma!} \frac{\partial^n}{\partial x_{ij}^\alpha \partial y_{ij}^\beta \partial z_{ij}^\gamma} \left[ x_{ij}^\alpha y_{ij}^\beta (-2z_{ij})^\gamma \varepsilon^n(r_{ij}) \psi(\mathbf{r}_1, \mathbf{r}_2, \ldots \mathbf{r}_N) \right]$$

$$= \frac{1}{n!}(-1)^{n-1} \left( \frac{1}{2} \right)^n \sum_{\alpha+\beta+\gamma=n} \frac{n!}{\alpha!\beta!\gamma!} \frac{\partial^n}{\partial x^\alpha \partial y^\beta \partial z^\gamma} \left[ x^\alpha y^\beta (-2z)^\gamma \varepsilon^n(r) \psi_{ij}(\mathbf{r}) \right]$$

, (39)

where $0 \leq \alpha, \beta, \gamma \leq n$. Therefore, the final expression for perturbation expansion in the case of uniaxial extension is:

$$g(\mathbf{r}) = g(r) + \sum_{n=1}^{\infty} \frac{1}{n!}(-1)^{n-1} \left( \frac{\lambda}{2} \right)^n \sum_{\alpha+\beta+\gamma=n} \frac{n!}{\alpha!\beta!\gamma!} \frac{\partial^n}{\partial x^\alpha \partial y^\beta \partial z^\gamma} \left[ x^\alpha y^\beta (-2z)^\gamma u^n(r) g(r) \right], \quad (40)$$

where $\varepsilon(r) = \lambda u(r)$.

## B. First- and Second-Order Expansions

We now proceed to the analysis of the first- and second-order expansion terms using the derived general expression [Eq. (40)]. For the $n = 1$ term, we have:

$$\frac{\lambda}{2}\left(x\frac{\partial}{\partial x} + y\frac{\partial}{\partial y} - 2z\frac{\partial}{\partial z}\right)[u(r)g(r)]$$

$$= \frac{1}{2}\left[(1 - 3\cos^2\theta)r\frac{\partial}{\partial r} + 3\sin\theta\cos\theta\frac{\partial}{\partial \theta}\right][\varepsilon(r)g(r)]. \quad (41)$$

$$= -\frac{1}{\sqrt{5}}r\frac{d}{dr}[\varepsilon(r)g(r)]Y_2^0(\theta,\phi)$$

Therefore,

$$g_2^0(r) = -\frac{1}{\sqrt{5}}r\frac{d}{dr}[\varepsilon(r)g(r)], \quad (42)$$

$$\varepsilon(r) = \varepsilon(D_0)\frac{g(D_0)}{g(r)} - \frac{\sqrt{5}}{g(r)}\int_{D_0}^{r}\frac{g_2^0(r')}{r'}dr'. \quad (43)$$

In the reciprocal space:

$$S_2^0(Q) = \frac{4\pi\rho}{\sqrt{5}}\int_0^\infty j_2(Qr)r^3\frac{d}{dr}[\varepsilon(r)g(r)]dr = \frac{4\pi\rho}{\sqrt{5}}\int_0^\infty \varepsilon(r)g(r)r\left[r\cos(Qr) - \frac{\sin(Qr)}{Q}\right]dr. \quad (44)$$

As in the case of shear flow, Eqs. (43) and (44) can be used to analyze the microscopic strain in the real and reciprocal spaces, respectively.

For the $n = 2$ term, the *effective* differential operator on $u^2(r)g(r)$ is:

$$\frac{\partial^2}{\partial x^2}x^2 + \frac{\partial^2}{\partial y^2}y^2 + 4\frac{\partial^2}{\partial z^2}z^2 + 2\frac{\partial^2}{\partial x\partial y}xy - 4\frac{\partial^2}{\partial x\partial z}xz - 4\frac{\partial^2}{\partial y\partial z}yz$$

$$= (3\cos^2\theta - 1)^2 r^2\frac{d^2}{dr^2} + (2 + 15\cos^2\theta - 9\cos^4\theta)r\frac{d}{dr} + 6 \quad (45)$$

Here, the differential operators associated with $\theta$ and $\phi$ are omitted, as $\varepsilon^2(r)g(r)$ depends only on $r$. Thus, the $n = 2$ term is:

$$\frac{1}{8}\left[(3\cos^2\theta - 1)^2 r^2 \frac{d^2}{dr^2} + (2 + 15\cos^2\theta - 9\cos^4\theta)r\frac{d}{dr} + 6\right]\left[\varepsilon^2(r)g(r)\right]. \tag{46}$$

Combining this result with the first-order term, one can show that the expansion up to the second order in $\lambda$ involves only three real spherical harmonics: $Y_0^0$, $Y_2^0$, and $Y_4^0$, with

$$g_0^0(r) - g(r) = \left(\frac{1}{10}r^2\frac{d^2}{dr^2} + \frac{13}{20}r\frac{d}{dr} + \frac{3}{4}\right)\left[\varepsilon^2(r)g(r)\right], \tag{47}$$

$$g_2^0(r) = \left(\frac{1}{7\sqrt{5}}r^2\frac{d^2}{dr^2} + \frac{17}{28\sqrt{5}}r\frac{d}{dr}\right)\left[\varepsilon^2(r)g(r)\right] - \frac{1}{\sqrt{5}}r\frac{d}{dr}[\varepsilon(r)g(r)], \tag{48}$$

$$g_4^0(r) = \left(\frac{3}{35}r^2\frac{d^2}{dr^2} - \frac{3}{35}r\frac{d}{dr}\right)\left[\varepsilon^2(r)g(r)\right]. \tag{49}$$

As in the case of shear flow, we can focus on $g_0^0(r)$ and integrate Eq. (47) to obtain an explicit expression for the microscopic strain:

$$\varepsilon^2(r) = \frac{c_1}{r^5 g(r)} + \frac{c_2}{r^{3/2}g(r)} - \frac{1}{r^5 g(r)}\int_{D_0}^r t^4 f(t)dt + \frac{1}{r^{3/2}g(r)}\int_{D_0}^r \sqrt{t}f(t)dt, \tag{50}$$

where $f(t) = \frac{20}{7}[g_0^0(t) - g(t)]$. In the reciprocal space, we have:

$$S_0^0(Q) - S(Q) = \frac{\pi\rho}{5}\int_0^\infty j_0(Qr)r^2\left(2r^2\frac{d^2}{dr^2} + 13r\frac{d}{dr} + 15\right)\left[\varepsilon^2(r)g(r)\right]dr. \tag{51}$$

Similarly, when $\varepsilon(r)$ is a constant, Eqs. (42), (44), (47), and (51) can be simplified:

$$\varepsilon = -\frac{\sqrt{5}g_2^0(r)}{r\frac{dg(r)}{dr}}, \tag{52}$$

$$\varepsilon = \frac{\sqrt{5}S_2^0(Q)}{Q\frac{dS(Q)}{dQ}}, \tag{53}$$

$$\varepsilon^2 = \frac{20\left[g_0^0(r) - g(r)\right]}{2r^2\frac{d^2g(r)}{dr^2} + 13r\frac{dg(r)}{dr} + 15g(r)}, \tag{54}$$

$$\varepsilon^2 = \frac{20\left[S_0^0(Q) - S(Q)\right]}{2Q^2\frac{d^2S(Q)}{dQ^2} + 3Q\frac{dS(Q)}{dQ}}, \tag{55}$$

where Eqs. (52) and (53) are the formulas used in previous studies [3, 19].

## IV. TESTS OF FORMULAS

### A. Justification of Perturbation Expansion Approach

In this section, we test the formulas developed in the preceding discussions. First, we would like to verify that it is indeed feasible to apply the perturbation expansion approach to analyze the anisotropic PDF or structure factor, i.e., higher order terms can be neglected when the deformation is small. For this purpose, we numerically study the single-chain structure factor of a Gaussian chain. Under affine shear deformation, the anisotropic $S(\mathbf{Q})$ is:

$$S(\mathbf{Q}) = \frac{2}{x^2}\left(e^{-x} + x - 1\right), \tag{56}$$

where $x = Q^2 R_g^2 \left(1 + \gamma \sin^2\theta \sin 2\phi + \gamma^2 \sin^2\theta \cos^2\phi\right)$, with $R_g$ being the equilibrium radius of gyration. In the isotropic case, i.e. $\gamma = 0$, Eq. (56) reduces to the well-known Debye function. The spherical harmonic expansion coefficients $S_l^m(Q)$ can be straightforwardly computed by carrying out weighing integrals with Eq. (19). The maximum relative errors of strain estimated from Eqs. (33) and (35) are defined as:

$$\Delta\gamma_2^{-2} = \max\left(\left|\gamma - \gamma_2^{-2}\right|/\gamma\right), \tag{57}$$

$$\Delta \gamma_0^0 = \max\left(\left|\gamma - \gamma_0^0\right|/\gamma\right), \tag{58}$$

where $\gamma_2^{-2}$ and $\gamma_0^0$ are:

$$\gamma_2^{-2} = \sqrt{15} S_2^{-2}(Q) \bigg/ \left[ Q \frac{dS(Q)}{dQ} \right], \tag{59}$$

$$\gamma_0^0 = \sqrt{30\left[S_0^0(Q) - S(Q)\right] \bigg/ \left[ Q^2 \frac{d^2 S(Q)}{dQ^2} + 4Q \frac{dS(Q)}{dQ} \right]}. \tag{60}$$

Additionally, the differences between the analytical model result and the perturbation expansion up to the second order in $\lambda$ for $S_2^{-2}(Q)$ and $S_0^0(Q)$ are

$$\Delta S_2^{-2}(Q) = S_2^{-2}(Q) - \frac{\gamma}{\sqrt{15}} Q \frac{dS(Q)}{dQ}, \tag{61}$$

$$\Delta S_0^0(Q) = S_0^0(Q) - S(Q) - \frac{\gamma^2}{30}\left[ Q^2 \frac{d^2 S(Q)}{dQ^2} + 4Q \frac{dS(Q)}{dQ} \right], \tag{62}$$

where $S_2^{-2}(Q)$ and $S_0^0(Q)$ are the exact values computed from Eq. (56).

Table 1 presents the relative errors of strain estimated by Eqs. (57) and (58), whereas Fig. 1(a) and (b) show how the difference between the analytical calculation and perturbation expansion varies with $Q$. It is evident that as expected the second-order formula has higher accuracy than the first-order one. For the strains investigated here, the relative error from the second-order equation is smaller than that from the first-order equation by approximately an order of magnitude. These results support the validity of our perturbation expansion approach, in which the high order terms can be neglected.

Table 1. Relative errors from perturbation expansion

|  | γ = 0.1 | γ = 0.2 | γ = 0.3 |
|---|---|---|---|
| $\Delta \gamma_2^{-2}$ | $3.4 \times 10^{-3}$ | $1.4 \times 10^{-2}$ | $3 \times 10^{-2}$ |

| $\Delta\gamma_0^0$ | $9.6 \times 10^{-4}$ | $3.8 \times 10^{-3}$ | $8.5 \times 10^{-3}$ |

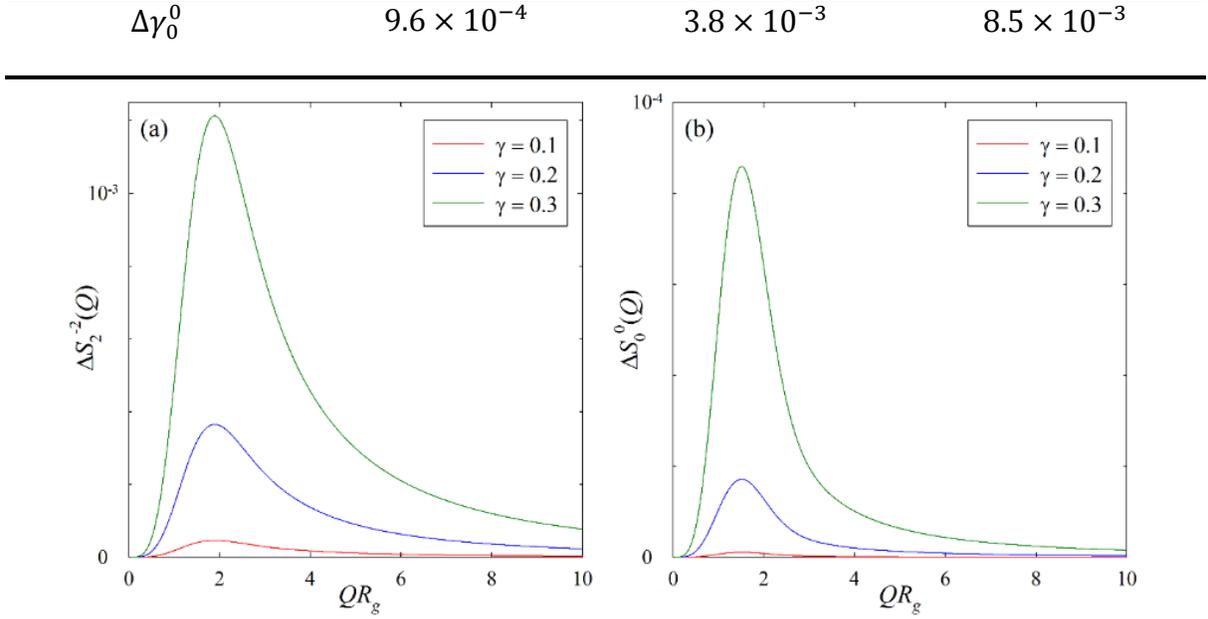

FIG. 1. The variation between the analytical formula [Eq. (56)] and the formulas of perturbation expansion [Eqs. (33, 35)] as a function of $QR_g$ for (a) $S_2^{-2}(Q)$ and (b) $S_0^0(Q)$ with different shear strains.

B. Evaluation of Microscopic Strain

To test the formulas for spatially varying strain, we have performed non-equilibrium molecular dynamics simulations of interacting particles under shear, using the LAMMPS software [31]. Our simulation box contains 16,000 particles at a fixed density of $\rho = 0.07843$ Å$^{-3}$. The pairwise interaction is described by the modified Johnson potential [32], which was developed for liquid iron. The SLLOD algorithm coupled with Nose-Hoover thermostat is applied to the system to simulate continuous shear [33]. Each time-step is set to be 1 fs. For the equilibrium PDF $g(r)$ and expansion coefficient $g_2^{-2}(r)$ shown in Fig. 2, the temperature is 1500 K and the shear rate is $1.3143 \times 10^{11}$ sec$^{-1}$. These data were fitted within $r \pm 1.3$ Å using Eqs. (22) and Eq. (37) to extract the microscopic strain $\gamma(r)$. The result in Fig. 3 is consistent with our assumption that Eq. (22) will reduce to Eq. (37) when the mechanical perturbation is small.

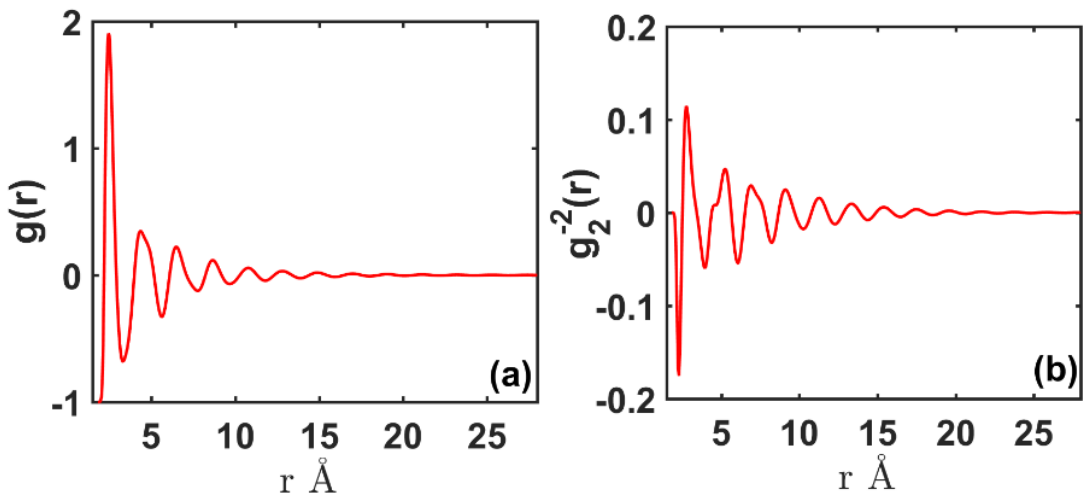

FIG. 2. The (a) $g(r)$ and (b) $g_2^{-2}(r)$ from molecular dynamics simulation using the LAMMPS software and SLLOD algorithm, where $N = 16,000$, $\rho = 0.07843 \text{ Å}^{-3}$, shear rate $1.3143 \times 10^{11}$ sec$^{-1}$, temperature 1500 K, time step 1 fs, and modified Johnson potential were used.

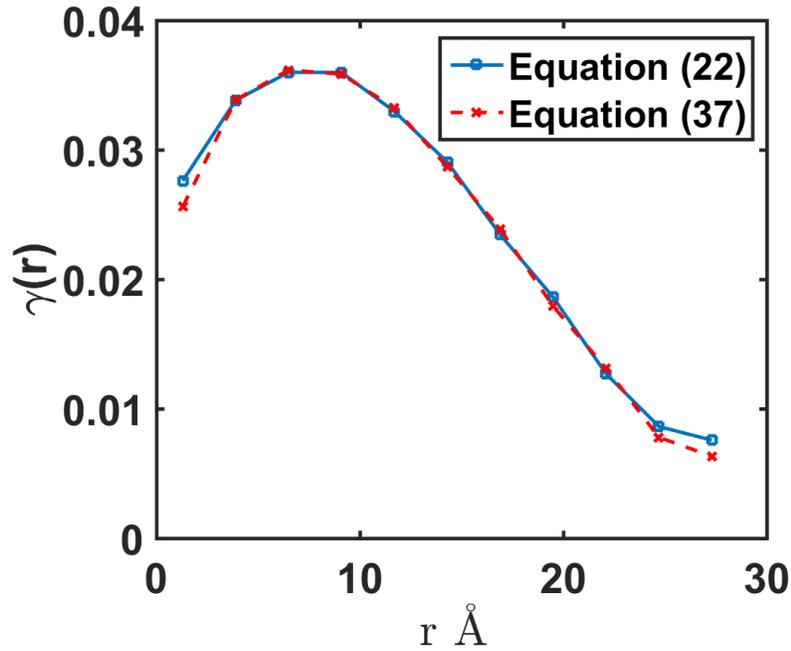

FIG. 3. The non-affine strain distribution as a function of $r$ predicted by Eq. (22) and Eq. (37), where the data points were obtained from molecular dynamics simulation using LAMMPS software and SLLOD algorithm.

## V. DISCUSSIONS AND CONCLUSIONS

In conclusion, equations for extracting microscopic strain in shear and uniaxial extension have been derived through the perturbation expansion of anisotropic PDF in terms of real spherical harmonics. The resulting formulas, presented in both real and reciprocal spaces, connect the anisotropic PDF and structure factor in the non-equilibrium state to the corresponding isotropic ones in the quiescent state. This phenomenological approach is independent of potential form, and can be applied to analysis of small-angle scattering experiments and computer simulations of materials under flow and deformation. Additionally, our formulas reduce to the affine ones as derived in Refs. [3, 4, 8, 11, 19, 20], and are equivalent to the non-affine strain equation found in Refs. [19, 20] with the condition: $d\ln\gamma(r)/dr \ll d\ln g(r)/dr$. We further demonstrate that the high order terms of real spherical harmonics cannot be ignored when the mechanical perturbation rises to a certain value. Unlike the situation for the zeroth and first order expansion, the isotropic terms $dg(r)/dr \neq dg_0^0(r)/dr$ and $dS(Q)/dQ \neq dS_0^0(Q)/dQ$ for the second expansion.

In the past, significant research effort has been devoted to the derivation of anisotropic PDF under flow and deformation, by assuming spatially homogeneous strain [1-13]. In some other studies [19, 20, 22, 23], expressions for the microscopic strain are defined by phenomenologically extending the affine equations [Eqs. (32) and (52)]:

$$\gamma(r) \equiv -\frac{\sqrt{15}g_2^{-2}(r)}{r\dfrac{dg_0^0(r)}{dr}},$$

$$\varepsilon(r) \equiv -\frac{\sqrt{5}g_2^0(r)}{r\dfrac{dg_0^0(r)}{dr}}.$$

These equations were subsequently used to fit the experimental $g_2^{-2}(r)$ or $g_2^0(r)$ data to connect the non-affine strain to the anisotropic PDF and inter-particle structure factor. However, such an approach lacks a rigorous mathematical foundation. In this work, by starting from the transformation of molecular displacement, we derive formulas for microscopic strain through the Irving-Kirkwood perturbation expansion method. Our method can be applied to analysis in both the real and the reciprocal spaces. For example, the strain distribution could be a key to distinguish the shear transformation zone and dynamical correlated region in the systems of metallic glass and colloid [22, 23].

We note that the traditional methods for extracting the microscopic strain rely on the derivative of $g(r)$ or $S(Q)$. This sometimes can be challenging, as good data statistics is required. Consequently, Eqs. (23) and (44) might be a better choice for the data analysis. Furthermore, Eqs. (31) and (51) permit characterization of the microscopic strain through $S_0^0(Q)$, instead of $S_2^{-2}(Q)$ (in the case of shear) or $S_2^0(Q)$ (in the case of uniaxial extension). As demonstrated by the numerical study of the affine model of Gaussian chains (Table 1), the strain estimation from $S_0^0(Q)$ has higher accuracy than the ones extracted from $S_2^{-2}(Q)$ or $S_2^0(Q)$. In the case of non-affine deformation, we also find that the data quality of $\gamma(r)$ evaluated by $g(r)$ is better than the one by $g_0^0(r)$. This stems from the fact that $dg_0^0(r)/dr$ [$dS_0^0(Q)/dQ$] is actually not equal to $dg(r)/dr$ [$dS(Q)/dQ$] for high order expansion even when structural difference is small (Fig. 1).

ACKNOWLEDGEMENT

This work was sponsored by the U.S. Department of Energy, Office of Science, Office of Basic Energy Sciences, Materials Sciences and Engineering Division. This research at the Spallation Neutron Source and the Center for Nanophase Materials Sciences of Oak Ridge National Laboratory was sponsored by the Scientific User Facilities Division, Office of Basic Energy Sciences, U.S. Department of Energy. Y.Y.W. thanks the support by the Laboratory Directed Research and Development Program of Oak Ridge National Laboratory, managed by UT

Battelle, LLC, for the U.S. Department of Energy. G.R.H. acknowledges the supports from the National Center for Theoretical Sciences, Ministry of Science and Technology in Taiwan (Project No. MOST 106-2119-M-007-019) and the Shull Wollan Center during his visit to Oak Ridge National Laboratory.

REFERENCES

[1] N. A. Clark and B. J. Ackerson, Phys. Rev. Lett. **44**, 1005 (1980).

[2] B. J. Ackerson and N. A. Clark, Physica A **118**, 221 (1983).

[3] Y. Suzuki, J. Haimovich, and T. Egami, Phys. Rev. B **35**, 2162 (1987).

[4] W. T. Ashurst and W. G. Hoover, Phys. Rev. A **11**, 658 (1975).

[5] S. Hess, Phys. Rev. A **22**, 2844 (1980).

[6] S. Hess and H. J. M. Hanley, Phys. Rev. A **25**, 1801(R) (1982).

[7] S. Hess and H. J. M. Hanley, Phys. Lett. A **98**, 35-38 (1983).

[8] D. J. Evans, H. J. M. Hanley, and S. Hess, Phys. Today **37**(1), 26 (1984).

[9] S. Hess, Int. J. Thermophys. **6**, 657 (1985).

[10] J. F. Schwarzl and S. Hess, Phys. Rev. A **33**, 4277 (1986).

[11] H. J. M. Hanley, J. C. Rainwater, and S. Hess, Phys. Rev. A **36**, 1795 (1987).

[12] H. J. M. Hanley and D. J. Evans, J. Chem. Phys. **76**, 3225 (1982).

[13] J. F. Morris and B. Katyal, Phys. Fluids **14**, 1920 (2002).

[14] P. Schall, D. A. Weitz, and F. Spaepen, Science **318**, 1895 (2007).

[15] V. Chikkadi and P. Schall, Phys. Rev. E **85**, 031402 (2012).

[16] F. Varnik, S. Mandal, V. Chikkadi, D. Denisov, P. Olsson, D. Vågberg, D. Raabe, and P. Schall, Phys. Rev. E **89**, 040301(R) (2014).

[17] A. Zaccone, P. Schall, and E. M. Terentjev, Phys. Rev. B **90**, 140203(R) (2014).

[18] S. Mandal, V. Chikkadi, B. Nienhuis, D. Raabe, P. Schall, and F. Varnik, Phys. Rev. E **88**, 022129 (2013).

[19] W. Dmowski, T. Iwashita, C.-P. Chuang, J. Almer, and T. Egami, Phys. Rev. Lett. **105**, 205502 (2010).

[20] T. Iwashita, and T. Egami, Phys. Rev. Lett. **108**, 196001 (2012).

[21] Z. Wang, C. N. Lam, W.-R. Chen, W. Wang, J. Liu, Y. Liu, L. Porcar, C. B. Stanley, Z. Zhao, K. Hong, and Y. Wang, Phys. Rev. X **7**, 031003 (2017).

[22] Z. Wang *et al*., submitted.

[23] T. Egami, Y. Tong, W. Dmowski., Metals **6**, 22 (2016).

[24] J. H. Irving and J. G. Kirkwood, J. Chem. Phys. **18**, 817 (1950).

[25] N. J. Wagner and W. B. Russel, Phys. Fluids A **2**, 491 (1990).

[26] N. J. Wagner and B. J. Ackerson, J. Chem. Phys. **97**, 1473 (1992).

[27] B. J. Maranzano and N. J. Wagner, J. Chem. Phys. **117**, 10291 (2002).

[28] A. K. Gurnon and N. J. Wagner, J. Fluid. Mech. **769**, 242 (2015).

[29] J.-P. Hansen and I. R. McDonald, *Theory of Simple Liquids* (Academic Press, London, 2013).

[30] Guan-Rong Huang, Yangyang Wang, Bin Wu, Zhe Wang, Changwoo Do, Gregory S. Smith, Wim Bras, Lionel Porcar, Péter Falus, and Wei-Ren Chen, Phys. Rev. E **96**, 022612 (2017).

[31] S. Plimpton, J. Comput. Phys. **117**, 1 (1995).

[32] D. Srolovitz, K. Maeda, V. Vitek, and T. Egami, Philos. Mag. A **44**, 847 (1981).

[33] D. J. Evans and G. P. Morriss, Phys. Rev. A. **30**, 1528 (1984).